\documentclass[a4paper]{jpconf} 
\usepackage{graphicx}
\usepackage{amssymb,amsmath}
\usepackage{textcomp}
\usepackage{subfigure}
\usepackage{cleveref}
\usepackage{braket}
\usepackage{subfigure}

\begin{document}

\title{Gauge fields renormalization groups and thermofractals}

\author{A. Deppman}
\address{Instituto de F\'{\i}sica, Rua do Mat\~ao 1371-Butant\~a, S\~ao Paulo-SP, CEP 05580-090, Brazil}
\ead{deppman@usp.br}

\author{E. Meg\'{\i}as}
\address{Departamento de F\'{\i}sica At\'omica, Molecular y Nuclear and Instituto Carlos I de F\'{\i}sica Te\'orica y Computacional, Universidad de Granada, Avenida de Fuente Nueva~s/n, 18071 Granada, Spain}

\author{D. P. Menezes}
\address{Departamento de F\'{\i}sica, CFM-Universidade Federal de Santa Catarina, Florian\'opolis, SC-CP. 476-CEP
88.040-900, Brazil}

\vspace{10pt}
\begin{indented}
\item[]March 2022
\end{indented}

\begin{abstract}
 The perturbative approach to QCD has shown to be limited, and the difficulties to obtain accurate calculations in the low-energy region seems to be insurmountable. A recent approach uses the fractal structures of Yang-Mills Field Theory to circumvent those difficulties, allowing for the determination of an analytic expression for the running coupling. The results obtained are in agreement with several experimental findings, and explain many of the observed phenomena at high-energy collisions. In this work, we address some of the conceptual aspects of the fractal approach, which are expressed in terms of the renormalization group equation and the self-energy corrections to the parton mass. We associate these well-known concepts with the origins of the fractal structure in the quantum field theory.
\end{abstract}



\section{Introduction}
In this paper we discuss the connections between fractals, Yang-Mills Fields and QCD, addressing the most central conceptual aspect of these connections, which appears in the scaling-properties of the field theory, the renormalization procedure, and the complex structure of the effective parton. A short review can be found in Ref.~\cite{Deppman:2020jzl}.

The data produced in high energy collisions have evidenced a robust feature of the multiparticle production dynamics: the power-law distributions of the particles energy or momenta, which is practically independent of the particle species and the collision energy  above $\sim~$1~TeV. The long-tail distributions were already noticed some decades ago, being a critical failure of the Hagedorn's Self-Consistent Thermodynamics and Chew and Frautisch's Bootstrap Model for the hadron structure~\cite{Hagedorn:1965st,Chew:1961ev}. See~\cite{Deppman:2017igr} for a short description of the problems faced by these approaches, and how the theory discussed here solves those problems.

A variety of explanations have been proposed, since then, to explain those exotic distributions. Initially, it was supposed that the large momentum sector would represent the exponential behaviour expected by the thermodynamics models, while the low momentum peak would be affected by a feed-up mechanism due to the particles decay after the freeze-out~\cite{Venugopalan:1992hy}. This approach went into discredit when collisions at higher energy allowed the production of a large number of more massive particles. Despite their longer and more diverse decaying chains, the observed distribution resulted to be similar to those of light particles, unveiling the universal character of the multiparticle production process. A second approach reverted the way as the long-tail was interpreted. In this case, the low momentum peak of the distributions represents the decay of a thermodynamically equilibrated system, while the high momentum region would be affected by hard-scattering that can be described by perturbative QCD (pQCD). The calculations with a few orders in pQCD give a reasonable result when compared with the long-tail region, but fails in describing the low-energy peak~\cite{Wong:2015mba}.

A third approach uses Tsallis Statistics~\cite{Tsallis:1987eu} to describe the thermodynamic properties of the hot and dense system produced at high energy collisions~\cite{Bediaga:1999hv,Beck:2000nz}. This approach can describe the entire range of momentum distribution in a single theoretical framework~\cite{Cleymans:2011in,Cleymans:2012ya, Wilk:2000yv,Wilk:2009nn}. When the Hagedorn's theory is generalized with the introduction of Tsallis Statistics, it can describe not only the high energy distributions but also the hadron mass spectrum, better than the original theory can do. The non-extensive self-consistent thermodynamics that results from this combined theory predicts that the Hagedorn Temperature and the entropic index, $q$, must be both constants at any collision energy above 1~TeV and for any particle species. Many works presented analyses of experimental data showing that those predictions are in good agreement with data~\cite{Azmi:2014dwa,Campos:2018tsb,Bhattacharyya:2020sua,Khandai:2013gva,Li:2013kca}, although some recent and more accurate analyses, based on high statistics data, indicate a small variations of $q$ with energy and particle species~\cite{Cleymans:2013rfq,Cleymans:2020ojr,Shen:2019zgi,Sharma:2020paa,Parvan:2019bga}. A comprehensive account of the subject can be found in Ref.~\cite{Biro:2017arf}

At least three different mechanisms were proposed to explain the emergence of the non-extensive statistics in HEP: the temperature fluctuations~\cite{Wilk:2009nn}, the small size of the system~\cite{Biro:2012fiy} and the fractal structure~\cite{Deppman:2016fxs}. All these mechanisms result in the same q-exponential distributions for the multiparticle production process, that can fit the data rather accurately. There are some reasons to believe that the three different mechanisms are indeed three different aspects of the same phenomenon. The fractal structure leads to temperature fluctuations that are the same used to show the emergence of Tsallis statistics. The small size description is also appropriate to explain the emergence of non-extensivity in the fractal structure, as will be discussed below.

The fundamental property of QCD that leads to the appearance of fractal structures is the scaling invariance of the vertex-functions, which is described by the Callan-Symanzik equation~\cite{Callan:1972uj,Symanzik1}. The scaling invariance is also one of the most important features of fractals. However, this property alone is not sufficient to characterize a system as a fractal, since these systems need to present a complex internal structure. Combined, the scaling property and the internal structure produce the major feature of fractals: the self-similarity. The vacuum polarization is an essential part of the interaction not only in QCD but also in QED and Yang-Mills fields in general, and the vacuum structure is an important component of partons interactions and the parton self-energy~\cite{Casher:1973uf}.

\section{Fractal structures in Yang-Mills fields}

\begin{figure}
\hspace{3.5cm} \begin{subfigure}{}
  \includegraphics[width=0.13\textwidth]{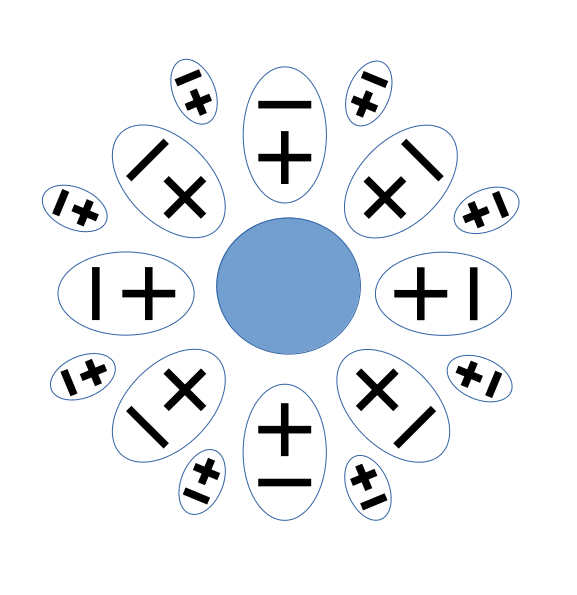}
 \end{subfigure}
\hspace{0.5cm} \begin{subfigure}{}
  \includegraphics[width=0.35\textwidth]{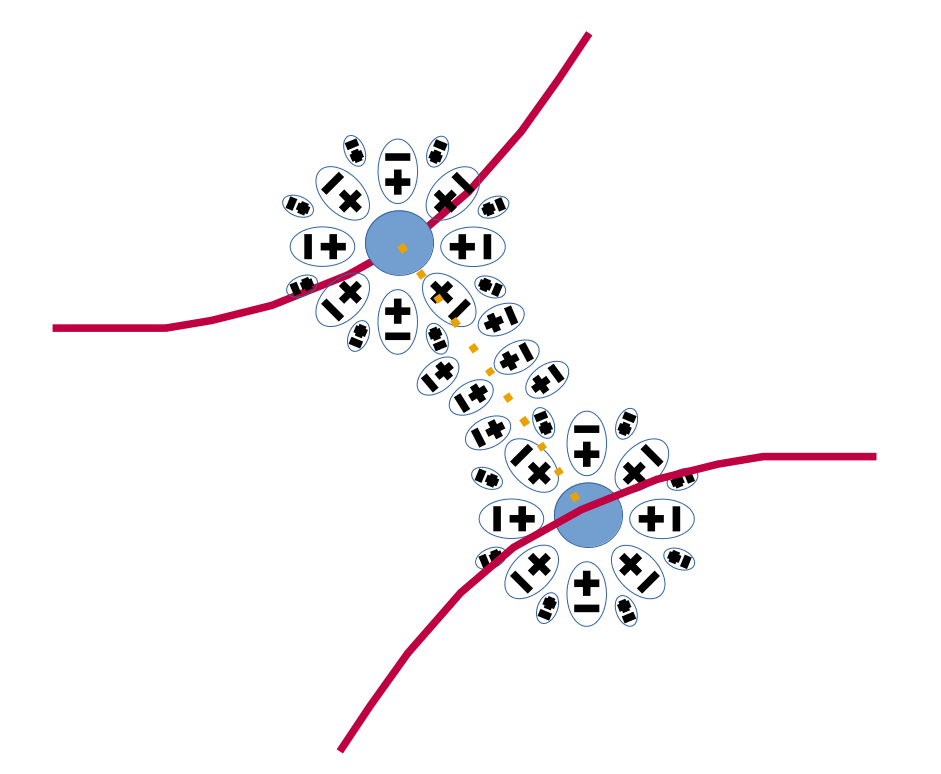}
 \end{subfigure}
\\
\indent  \hspace{4.4cm} {\bf (a)}  \hspace{4.2cm}  {\bf (b)}
 \caption{Pictorial representation of the effective parton and their  interaction. (a) The vacuum polarization represented as the internal structure of the effective parton. (b) In the effective parton interaction the vacuum structure participates in a complex way. }
 \label{fig:fractalparton}
\end{figure}

The partonic dynamics is described, in terms of the Dyson-Schwinger expansion~\cite{Dyson:1949ha}, as
\begin{equation}
 \ket{\Psi_n} =  \sum_{\{n\}} (-i)^n \int dt_1 \dots dt_n \ket{\Psi}\bra{\Psi} e^{-iH_o(t_n-t_{n-1})} g  \dots e^{-iH_o (t_1-t_o)}  \ket{\Psi_o}, \label{eqn:elDS}
\end{equation} 
where the $n$th order quantum state $\ket{\Psi_n}$ is described in terms of the states $\ket{\Psi}$, that represent the free elementary parton, with an initial state, $\ket{\Psi_o}$. In the perturbative expansion of strongly interacting fields, the Hamiltonian has two parts, $H_o$ that represents the self-energy contribution, and $H_I=1+g\, dt$, that is the perturbative contribution of the proper-vertex interactions. A summation over all possible states is implicit in the terms $ \ket{\Psi}\bra{\Psi}$.

The effective parton, which includes the self-energy interaction in the propagation of the elementary parton \cite{Dyson:1949ha,Dyson:1956zza}, has states given by
\begin{equation}
 \ket{\Psi_e(t)}=e^{-iH_o (t-t_o)}  \ket{\Psi} \,. \label{eqn:eparton}
\end{equation}
The Dyson-Schwinger expansion can be written in terms of the effective parton states as
\begin{equation}
 \ket{\Psi_{e}(t_n)} =  \sum_{\{n\}} (-i)^n \int dt_1 \dots dt_n \ket{\Psi_e(t_n)}\bra{\Psi_e(t_n)} g \ket{\Psi_e(t_{n-1})}  \dots  g \ket{\Psi_e(t_o)}. \label{eqn:efDS}
\end{equation} 
Observe that $g$ represents proper-interaction vertices for the effective partons. As~\Cref{eqn:eparton} shows, the effective parton is a complex object that carries in its wave-function all the properties of its interaction with the vacuum.
The structure of the effective parton is depicted in~\Cref{fig:fractalparton}(a), where the vacuum polarization is represented by the $+$ and $-$ signs surrounding the elementary parton, represented by the central circle. In this sense we say that the effective parton has an internal structure. The  complexity of this structure can be evaluated by the number of Feynman graphs necessary to describe the self-interaction contributions even in low-orders of calculation can be appreciated by simply inspecting the figures in Ref.~\cite{Borowka:2018anu}. 

The proper-vertex interaction is still more complex, as can be observed in~\Cref{fig:fractalparton}(b). The interaction is mediated by another parton (boson)  which has its own self-energy contributions. The detailed description of all possible configurations is a huge challenge to perturbative QCD. The present situation indicates that the perturbative-QCD approach will not be able to provide an accurate calculation of the running-constant at low energies with a reasonable number of loops in the calculations.

The physical parton that can be experimentally observed is the effective parton, hereby, we only have access to the effective coupling, $g$. The Callan-Symanzik equation is the renormalization-group equation for the QCD, and applies to the effective parton and the effective coupling. It is written as
\begin{equation}
  \left[M\frac{\partial}{\partial M}  + \beta_g \frac{\partial}{\partial \bar{g}} + \gamma \right]\Gamma=0 \,, \label{CallanSymanzik}
\end{equation}
where $M$ is the scale parameter, and $\gamma$ is related to the form the effective parton fields scale. This equation establishes the scaling symmetry of the effective parton. Therefore, we get the two necessary conditions to fulfill the self-similarity of the physical parton: the scaling-free property and the internal structure. The work developed in~\cite{Deppman:2019yno} used these properties of the partons in any Yang-Mills field theory to associate the thermodynamical properties of those fields with the properties of thermofractals. A short review on the subject can be found in Ref.~\cite{Deppman:2020jzl}.


There are two types of thermofractals~\cite{Deppman:2016fxs}, and each one presents the following properties~\cite{Deppman:2019yno}:
\begin{enumerate}
  \item It is a thermodynamics system with a complex structure with a number $N$ of components that present the same properties as the parent system. It has a  total energy $U=E+K$, where $E$ is the internal energy of its components and $K$ is their total kinetic energy
  \item In the case of the thermofractal type-I, the internal energy, $E$, and the kinetic energy, $K$, of the components of the  system are such that the ratio $\varepsilon/\lambda=E/K$ follows a distribution $\tilde{P}(\varepsilon)$. In the case of the thermofractal type-II, it is the ratio $\varepsilon/\lambda=E/U$ that follows a distribution $\tilde{P}(\varepsilon)$. Here, $\lambda$ is a scaling parameter and $\varepsilon$ is the energy of the components of the thermofractal.
  \item At some level of the internal structure, the fluctuations of the internal energy of the system are small enough to be disregarded. Then their internal energy can be regarded as constant.
\end{enumerate}
The systems that present these properties must satisfy a recursive equation for the probability density, $\tilde{P}(\varepsilon)$, for which the only solution is~\cite{Deppman:2019yno,Deppman:2016fxs}
\begin{equation}
 \tilde{P}(\varepsilon) = \left[1\pm(q-1)\frac{\varepsilon}{\lambda}\right]^{\frac{\mp 1}{q-1}} \,, \label{eqn:qexps}
\end{equation}
where $q$ is related to $N$, and the exponent $1/(q-1)$ gives the number of degrees of freedom of the interacting system. Due to the property 3 of the thermofractals, at some value of $\lambda$ the scaling-symmetry is broken. Fixing $\lambda$ to this value transforms the function in~\Cref{eqn:qexps} in a q-exponential without (sign $+$ in the argument) and with (sign $-$ in the argument) cut-off. These functions are related to the Tsallis Statistics, and here they represent the connection between the thermofractals and the non-extensive statistics. The Yang-Mills fields form fractals of type-II~\cite{Deppman:2019yno}.

The fractal structure of Yang-Mills Fields were already investigated numerically~\cite{Wellner:1994cb,Wellner:1992rf}. The predictions of non-extensive thermodynamics theory discussed here were compared to Lattice QCD results, showing a good agreement~\cite{Deppman:2012qt}.

Notice that the second property of the thermofractals expresses the self-similarity of the thermofractals, and this is a fundamental aspect to obtain the recursive relation that leads to Tsallis Statistics. The q-exponential function describes how the energy and momenta of particles are shared at every vertex, and therefore is associated to the effective coupling by~\cite{Deppman:2019yno}
\begin{equation}
 g(\varepsilon) = G_o\prod_{i=1}^{2} \left[1 + (q-1)\frac{\varepsilon_i}{\lambda}\right]^{\frac{-1}{q-1}} \,, \label{eqn:effectivecoupling}
\end{equation}
where $i$ indicates the two particles in the final state at each vertex and $G_o$ is the overall strength of the interaction. This coupling satisfies the Callan-Symanzik Equation, and the connection between Tsallis distributions and the renormalization group equation were already investigated in~\cite{Deppman:2017rtf}.

An essential aspect of these systems is the fact that the number of degrees of freedom, that is given by the exponent of the functions in~\Cref{eqn:effectivecoupling}, is independent of the size of the system, that is given by the ratio $\varepsilon/\lambda$. In the case of systems following the Boltzmann-Gibbs Statistics, the number of degrees of freedom is connected to the size of the system. For an ideal gas, for instance, the number of degrees of freedom is proportional to the number of particles, and increases to infinity in the thermodynamical limit. In the case of Yang-Mills fields, the parameter $q$ can be calculated by using the fundamental parameters of the quantum field theory. For QCD, it was shown that~\cite{Deppman:2019yno,Deppman:2020gbu} $(q-1)^{-1}=(1/3)[11 n_c- (4/2) n_f]$
where $n_c$ is the number of colours and $n_f$ the number of flavours. The theoretical value obtained  for $q$ is in good agreement with those found in experimental data analyzes. A short review on the subject can be found in Ref.~\cite{Deppman:2020jzl}.

\section{Discussion}

The perturbative approach to Yang-Mills field theories present a complex internal structure that we associated with the self-energy interactions. Since the effective partons field must satisfy the renormalization group equation, they present also a scale-free symmetry. We argued that the combination of these two properties leads to the self-similarity of the effective parton structure. This is the main reason behind the formation of the fractals structures.

The fractal structures lead to recursive equations. The solution to those equations leads to q-exponential functions. From the statistical point of view, the structure is associated with thermofractals~\cite{Deppman:2016fxs}. This kind of system was introduced to explain the fractal origin of the non-extensive statistics described in terms of the Tsallis entropy. Therefore, we show that Yang-Mills systems are described thermodynamically by the Tsallis thermodynamics~\cite{Curado:1991jc,Tsallis:1998ws,Megias:2015fra}. In Ref.~\cite{Deppman:2019yno} the q-exponential functions of the partonic energy were considered as terms of the effective coupling in the interaction among the effective partons. Thus, even for QCD, the effective coupling is obtained non-perturbatively as an analytical expression. The self-similarity has been observed in high-energy collisions~\cite{Wilk:2013jsa,Zborovsky:2006nh,Moriggi:2020zbv}.

From the three parameters in the effective coupling, namely $q$, $\lambda$ and $G_o$, the first one is determined from the field theory parameters, $\lambda$ is the point where the scale symmetry is broken, and only $G_o$ remains to be determined. It represents the overall strength of the interaction, corresponding to the coupling constant at collision energies close to zero, and can be found by fittings to the experimental data. For QCD, the value for $q$ found theoretically is in good agreement with the value found in experimental data analyses~\cite{Cleymans:2011in,Sena:2012ds,Marques:2012px,Marques:2015mwa}.

Besides providing the effective coupling and the value for $q$, the theory also explains why Tsallis statistics must be used to describe the distributions obtained in high-energy collisions, since the use of the coupling constant in the form given in~\Cref{eqn:effectivecoupling} results in vertex functions corresponding to Tsallis distributions. It can be interesting to investigate if the subtle deviations of the $q$-exponential distribution~\cite{Cleymans:2020ojr,Wilk:2015pva} might indicate effects related to the finite quark mass of some flavours, that were not considered in the theoretical approach.

It is also possible to use the theory to calculate the fractal dimension.  This dimension can be accessed experimentally by the use of intermittency analysis~\cite{Bialas:1988wc,Hwa:1987dq,Dremin:1993ee}. The calculated value~\cite{Deppman:2016fxs}, $D=0.69$, is in good agreement with the experimental data analysis~\cite{Sarma:2019teo,Xie:2013rwa,Sarkisian:2000ux,Chen:1998ak,Rasool:2015ria}. A consequence of the fractal dimension and of the non-extensivity, is that the multiplicity of particles produced in high-energy collisions follows a power-law function of the energy~\cite{Deppman:2019yno}. The exponent predicted by the theory is in accordance with the one found in experimental data analysis~\cite{Sarkisyan:2015gca}.

The theory discussed here reconciles the Hagedorn Self-Consistent Thermodynamics~\cite{Hagedorn:1965st} with the high-energy experiments. This happens because, as we observed in the present work, the effective parton is a thermofractal of type-II~\cite{Deppman:2016fxs}, so it must be described by Tsallis Statistics~\cite{Tsallis:1987eu} and its associated thermodynamics~\cite{Curado:1991jc,Tsallis:1998ws}. The non-extensive self-consistent thermodynamics~\cite{Deppman:2016fxs} that results from the generalization of the Hagedorn's principle can describe the high-energy distributions, and predicts a constant value for $q$. But it also gives a new formula for the hadron mass-spectrum, which can describe the observed hadronic states even at masses as low as the pion mass~\cite{Marques:2012px}. This is an indication that hadrons at the normal conditions can be described by the same theory that describes the hot and dense medium created at high energy collisions. This aspect opens the possibility~\cite{Tan:2019zyw} to apply the theoretical approach discussed here to other systems, as hadron states~\cite{Andrade:2019dgy,Isayev:2018hzq}, cosmic rays~\cite{Yalcin:2017fvl} and neutron-stars~\cite{Menezes:2014wqa,Cardoso:2017pbu,Salako:2020tno}.

\section{Conclusions}

We discussed the fractal structures present in Yang-Mills fields, analyzing in details the concept of effective parton to show that it presents all the necessary features of a fractal. The self-similarity of the effective partons is the main characteristic that leads to recursive relations that allows for obtaining the running-coupling of QCD at the non-perturbative regimes.

We argue that the introduction of the proper-vertex in the perturbative approach to quantum fields calculation, that separates the self-energy contributions from the interaction mechanism, and the renormalization theory that determines the scaling properties of the effective parton, are the determinant features for the formations of the fractal structures.

A result of the introduction of the fractal methods in the quantum field theory is the possibility to determine the running-constant by an analytical formula with one single free parameter. The Tsallis distributions observed in high-energy collisions emerges naturally from the present approach. This theory opens the possibility to investigate several aspects of high-energy collisions and of hadrons. The fractal dimension, the non-extensive behaviour, and the self-similarity can be tested experimentally.

Several possible applications of the theory are mentioned, some of them already with interesting results.

\section{Acknowledgments} 

A D and D P M are partially supported by the Conselho Nacional de Desenvolvimento Cient\'{\i}fico e Tecnol\'ogico (CNPq-Brazil) and by Project INCT-FNA Proc. No. 464 898/2014-5. A D is partially supported by FAPESP under grant 2016/17612-7. The work of E M is supported by the project PID2020-114767GB-I00 financed by MCIN/AEI/10.13039/501100011033, by the FEDER/Junta de Andaluc\'{\i}a-Consejer\'{\i}a de Econom\'{\i}a y Conocimiento 2014-2020 Operational Programme under Grant A-FQM-178-UGR18, by Junta de Andaluc\'{\i}a under Grant FQM-225, and by the Consejer\'{\i}a de Conocimiento, Investigaci\'on y Universidad of the Junta de Andaluc\'{\i}a and European Regional Development Fund (ERDF) under Grant SOMM17/6105/UGR. The research of E M is also supported by the Ram\'on y Cajal Program of the Spanish MCIN under Grant RYC-2016-20678.

\vspace{1cm}

\section*{References}

\bibliographystyle{unsrt}
\bibliography{Bibliography_SelfConsistentHadron,Bibliography_RenormalizationYMF,Bibliography_TsallisHEP,Bibliography_ScalingHEP,Bibliography_OurPapers_Thermofractals,Bibliography_HadronResonanceGasModel,Bibliography_IntermittencyHEP,Bibliography_Tsallis,Bibliography_FractalHadrons}

\end{document}